\documentclass[prd,nofootinbib,preprint,showpacs,showkeys,superscriptaddress]{revtex4}

\usepackage[english]{babel}
\usepackage{amsmath,amssymb}
\usepackage[dvips]{graphicx}
\usepackage{ifthen,array}
\usepackage[sort&compress]{natbib}

\allowdisplaybreaks

\newcommand{\eg} {{\it e.g.}}

\newcommand{\CL}   {C.L.}
\newcommand{\dof}  {d.o.f.}
\newcommand{\eVq}  {\text{eV}^2}
\newcommand{\Sol}  {\textsc{sol}}

\newcommand{\Atm}  {\textsc{atm}}

\newcommand{\Dms}  {\Delta m^2_\Sol}
\newcommand{\Dma}  {\Delta m^2_\Atm}

\newcommand{\EtAl}  {{\it et al.\/}}

\newcommand{\flux}[2][]{\ensuremath{\ifthenelse{\equal{#1}{}}{}{^{#1}\!}\mathit{#2}}}

\newcommand{\AddrAHEP}{%
  Instituto de F\'{\i}sica Corpuscular --
  C.S.I.C./Universitat de Val{\`e}ncia \\
  Edificio Institutos de Paterna, Apt 22085,
  E--46071 Valencia, Spain}

\newcommand{\AddrTUM}{%
  Theoretische Physik, Physik Department, 
  Technische Universit{\"a}t M{\"u}nchen,
  James--Franck--Strasse, D--85748 Garching, Germany}

\begin{document}

\preprint{IFIC/03-42\hspace{1cm}TUM-HEP-528/03}

\vspace*{2cm} \title{Status of three--neutrino oscillations
  after the SNO--salt data}

\author{M.~Maltoni} \email{maltoni@ific.uv.es}
\affiliation{\AddrAHEP}

\author{T.~Schwetz} \email{schwetz@ph.tum.de}
\affiliation{\AddrTUM}

\author{M.~A.~T{\'o}rtola} \email{mariam@ific.uv.es}
\author{J.~W.~F.~Valle} \email{valle@ific.uv.es}
\affiliation{\AddrAHEP}

\begin{abstract}
  We perform a global analysis of neutrino oscillation data in the
  framework of three neutrinos, including the recent improved
  measurement of the neutral current events at SNO.  In addition to
  all current solar neutrino data we take into account the reactor
  neutrino data from KamLAND and CHOOZ, the atmospheric neutrino data
  from Super--Kamiokande and MACRO, as well as the first spectral data
  from the K2K long baseline accelerator experiment. The up-to-date
  best fit values and allowed ranges of the three--flavour oscillation
  parameters are determined from these data. Furthermore, we discuss
  in detail the status of the small parameters $\alpha \equiv
  \Dms/\Dma$ and $\sin^2\theta_{13}$, which fix the possible strength
  of CP violating effects in neutrino oscillations. 
\end{abstract}


\maketitle

\section{Introduction}

Recently, the Sudbury Neutrino Observatory (SNO)
experiment~\cite{ahmad:2003} has released an improved measurement with
enhanced neutral current sensitivity due to neutron capture on salt,
which has been added to the heavy water in the SNO detector.  This
adds precious information to the large amount of data on neutrino
oscillations published in the last few years. Thanks to this growing
body of data a rather clear picture of the neutrino sector is starting
to emerge. In particular, the results of the KamLAND reactor
experiment~\cite{eguchi:2002dm} have played an important role in
confirming that the disappearance of solar electron
neutrinos~\cite{cleveland:1998nv,davis:1994jw,abdurashitov:2002nt,abdurashitov:1999zd,hampel:1998xg,altmann:2000ft,cattadori:2002rd,fukuda:2002pe,ahmad:2002jz,ahmad:2002ka,ahmad:2001an},
the long--standing solar neutrino problem, is mainly due to
oscillations and not to other types of neutrino
conversions~\cite{guzzo:2001mi,barranco:2002te}.  Moreover, KamLAND
has pinned down that the oscillation solution to the solar neutrino
problem is the large mixing angle MSW solution
(LMA--MSW)~\cite{maltoni:2002aw,Bahcall:2002ij,fogli:2002au,deHolanda:2002iv}
characterized by the presence of matter
effects~\cite{wolfenstein:1978ue,mikheev:1985gs}.  On the other hand,
experiments with atmospheric
neutrinos~\cite{fukuda:1994mc,becker-szendy:1995vr,allison:1999ms,shiozawa:2002,fukuda:1998mi,surdo:2002rk}
show strong evidence in favour of $\nu_\mu\to\nu_\tau$ oscillations,
in agreement with the first data from the K2K accelerator
experiment~\cite{ahn:2002up}. Together with the non--observation of
oscillations in reactor experiments at a baseline of about
1~km~\cite{apollonio:1999ae,boehm:2001ik}, and the strong rejection
against transitions involving sterile states~\cite{maltoni:2002ni},
these positive evidences can be very naturally accounted for within a
three-neutrino framework. The large and nearly maximal mixing angles
indicated by the solar and atmospheric neutrino data samples,
respectively, come as a surprise for particle physics, as it contrasts
with the small angles characterizing the quark sector.
 
In this work we report on the implications of the new SNO--salt
measurement on the determination of three--neutrino oscillation
parameters. In Sec.~\ref{sec:solar+kaml} we investigate the impact of
the new data on the solar neutrino parameters by performing a fit to
solar data and the KamLAND reactor experiment in a simple
two--neutrino framework.  In Sec.~\ref{sec:3nu} we present the results
of the general three--neutrino fit to the global data from solar,
atmospheric, reactor and accelerator neutrino experiments.  After
fixing our notation in Subsec.~\ref{sec:notation} we investigate the
impact of three--flavour effects on solar neutrino parameters
(Subsec.~\ref{sec:sol3nu}) and we give the current best fit points and
allowed ranges for all three--flavour oscillation parameters
(Subsec.~\ref{sec:global}).  Furthermore, in
Subsec.~\ref{sec:subleading} we discuss in detail the constraints on
the small parameters $\sin^2\theta_{13}$ and the ratio $\Dms/\Dma$,
which are the crucial parameters governing the possible strength of CP
violating effects in neutrino oscillations. A summary of the
constraints on three--neutrino oscillation parameters from all current
experiments is given in Sec.~\ref{sec:conclusions}.

\begin{figure}[t] \centering
 \includegraphics[width=0.5\linewidth]{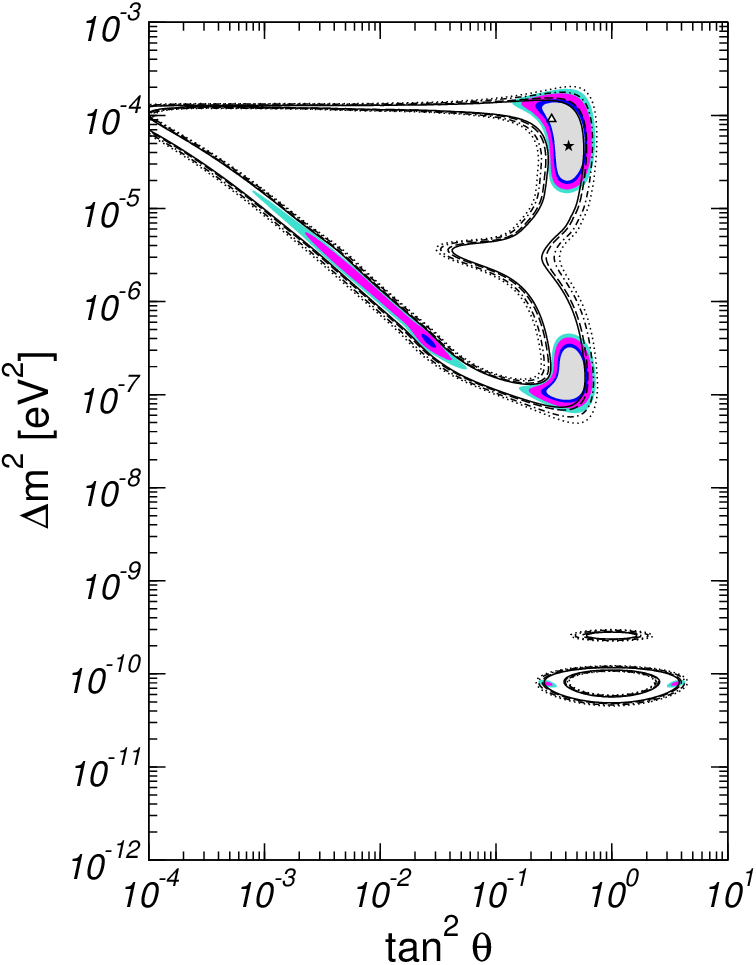}
    \caption{\label{fig:snototal} Allowed regions of
      $\sin^2\theta_\Sol$ and $\Dms$ at 90\%, 95\%, 99\% and 3$\sigma$
      \CL\ for 2 \dof\ from SNO--salt data~\cite{ahmad:2003} only
      (lines) and total SNO
      data~\cite{ahmad:2003,ahmad:2002jz,ahmad:2002ka} (colored
      regions).}
\end{figure}

\section{Two--neutrino analysis of solar and KamLAND data}
\label{sec:solar+kaml}

In this section we investigate the impact of the new SNO--salt data on
the determination of the solar neutrino parameters $\sin^2\theta_\Sol$
and $\Dms$ in a two--neutrino framework.  Before presenting the global
analysis of all solar data it is instructive to determine the
restrictions on solar neutrino oscillation parameters that follow from
SNO data only.  The regions delimited by the lines in
Fig.~\ref{fig:snototal} are obtained only from the new SNO--salt data
in the form of the neutral current (NC), charged current (CC) and
elastic scattering (ES) fluxes as reported in Ref.~\cite{ahmad:2003}.
To include these data we follow the prescription given in
Ref.~\cite{ahmad:2003a}, taking carefully into account of correlations
and systematic effects. The colored regions of
Fig.~\ref{fig:snototal} result from the total SNO data including the
new salt measurements~\cite{ahmad:2003}, as well as the 2002 spectral
day/night data~\cite{ahmad:2002jz,ahmad:2002ka}. Following
Ref.~\cite{ahmad:2003a} the 2002 and 2003 data are treated
uncorrelated.  For details of our SNO spectral analysis see
Ref.~\cite{maltoni:2002ni}.  One sees that Fig.~\ref{fig:snototal} is
in good agreement with the result obtained in Ref.~\cite{ahmad:2003}.
This figure shows also that other experiments play an important role
in ruling out low--mass solutions, such as the LOW solution, the small
mixing angle non--adiabatic branch of the MSW plot, as well as the
vacuum type solutions, still present here at the 99\% \CL.

With this calibration at hand, we now turn to the full two--neutrino
data analysis of the global set of solar neutrino data. In addition to
the SNO data we take into account the rates of the chlorine experiment
at the Homestake mine~\cite{cleveland:1998nv,davis:1994jw} ($2.56 \pm
0.16 \pm 0.16$~SNU), the most up-to-date results~\cite{taup} of the
gallium experiments
SAGE~\cite{abdurashitov:2002nt,abdurashitov:1999zd}
($69.1~^{+4.3}_{-4.2}~^{+3.8}_{-3.4}$~SNU) and
GALLEX/GNO~\cite{cattadori:2002rd,hampel:1998xg,altmann:2000ft} ($69.3
\pm 4.1 \pm 3.6$), as well as the 1496--day Super--Kamiokande data
sample~\cite{fukuda:2002pe} in the form of 44 bins (8 energy bins, 6
of which are further divided into 7 zenith angle bins). The analysis
methods used here are similar to the ones described in
Refs.~\cite{gonzalez-garcia:2000sq,maltoni:2002ni} and references
therein, with the exception that in the current work we use the
so-called pull-approach for the $\chi^2$ calculation. As described in
Ref.~\cite{Fogli:2002pt}, each systematic uncertainty is included by
introducing a new parameter in the fit and adding a penalty function
to the $\chi^2$. However, the method described in
Ref.~\cite{Fogli:2002pt} is extended in two respects. First, it is
generalized to the case of correlated statistical
errors~\cite{Balantekin:2003jm}, as necessary to treat the SNO--salt
data. Second, in the calculation of the total $\chi^2$ we use the
exact relation between the theoretical predictions and the pulls
associated to the solar neutrino fluxes, rather than keeping only the
terms up to first order. This is particularly relevant for the case of
the solar $^8$B flux, which is constrained by the new SNO data with an
accuracy better than the prediction of the Standard Solar Model
(SSM)~\cite{bahcall:1998wm}. In our approach it is possible to take
into account on the same footing both the SSM boron flux prediction
and the SNO NC measurement, without pre-selecting one particular
value.  In this way {\it the fit itself} can choose the best
compromise between the SNO NC data and the SSM value.

\begin{figure}[t] \centering
\includegraphics[width=.6\linewidth]{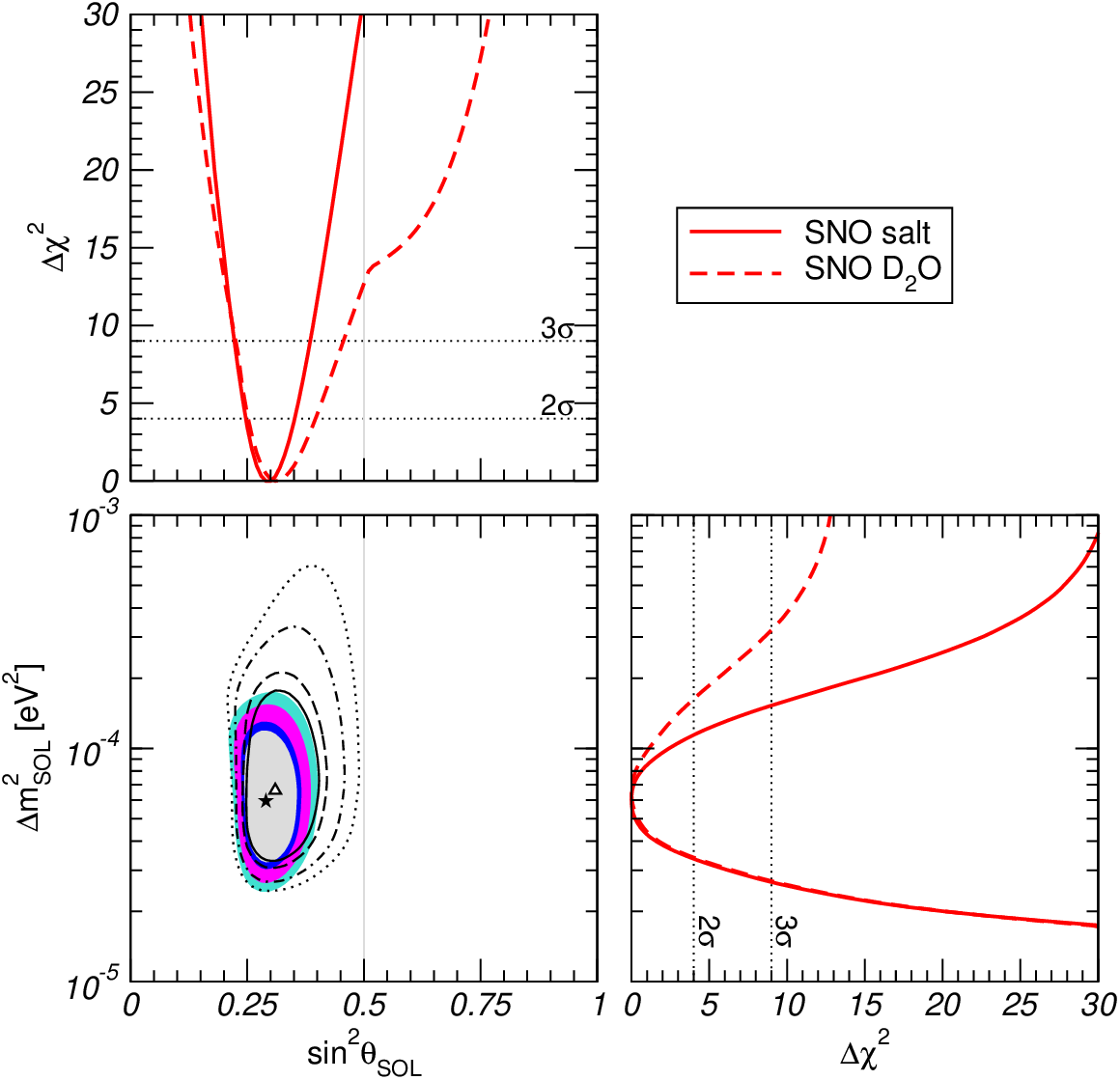}
  \caption{\label{fig:sol-region}%
      Projections of the allowed regions from all solar neutrino data
      at 90\%, 95\%, 99\%, and 3$\sigma$ \CL\ for 2 \dof\ onto the
      plane of $\sin^2\theta_\Sol$ and $\Dms$ before (lines) and
      after (colored regions) the inclusion of the SNO--salt data.
      Also shown is $\Delta \chi^2$ as a function of
      $\sin^2\theta_\Sol$ and $\Dms$, minimized with respect to the 
      un-displayed parameter.} 
\end{figure}

In Fig.~\ref{fig:sol-region} we compare the allowed regions for the
oscillation parameters before and after the new SNO--salt data.  
One finds that especially the upper part of the LMA--MSW region and
large mixing angles are strongly constrained by the new data. This
follows mainly from the rather small measured value of the CC/NC ratio
of $0.306\pm0.026\pm0.024$~\cite{ahmad:2003}, since this observable
increases when moving to larger values of $\Dms$ and/or
$\sin^2\theta_\Sol$ (see, \eg, Ref.~\cite{deHolanda:2002iv}). The best
fit values for the parameters are
\begin{equation}\label{eq:bfp-solar}
\sin^2\theta_\Sol = 0.29 \,,\qquad
\Dms = 6.0 \times 10^{-5}~\eVq \qquad\text{(solar data)}.
\end{equation}
The new data show a much stronger rejection against maximal solar
mixing: from the $\Delta \chi^2$ projected onto the
$\sin^2\theta_\Sol$ axis shown in Fig.~\ref{fig:sol-region} we find
that $\sin^2\theta_\Sol = 0.5$ is excluded at more than 5$\sigma$,
ruling out all bi--maximal models of neutrino
mass~\cite{pakvasa:2003zv}.

\begin{figure}[t] \centering
\includegraphics[width=.6\linewidth]{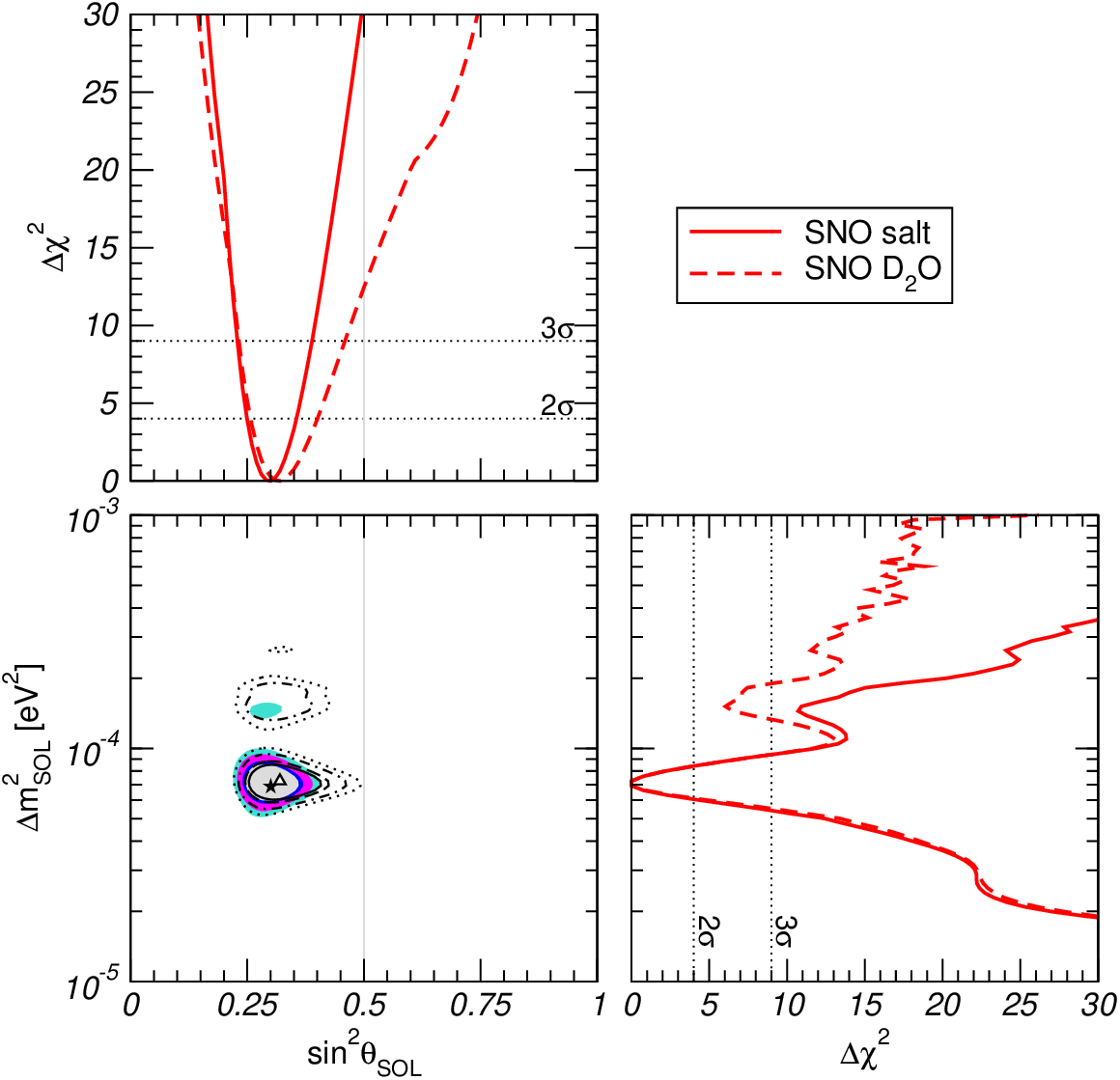}
    \caption{\label{fig:sol+kaml-region}%
      Projections of the allowed regions from all solar neutrino and
      KamLAND data
      at 90\%, 95\%, 99\%, and 3$\sigma$ \CL\ for 2 \dof\ onto the
      plane of $\sin^2\theta_\Sol$ and $\Dms$ before (lines) and
      after (colored regions) the inclusion of the SNO--salt data.
      Also shown is $\Delta \chi^2$ as a function of
      $\sin^2\theta_\Sol$ and $\Dms$, minimized with respect to the 
      un-displayed parameter.} 
\end{figure}

We now turn to the combination of solar data with data from the
KamLAND reactor experiment~\cite{eguchi:2002dm}.  To this aim we use
the event--by--event likelihood analysis for the KamLAND data as
described in Ref.~\cite{schwetz:2003se}, which gives stronger
constraints than a $\chi^2$--fit based on energy binned data.  In
Fig.~\ref{fig:sol+kaml-region} we show the results of the combined
analysis. Comparing post and pre--SNO--salt results (see, \eg,
Refs.~\cite{maltoni:2002aw,Bahcall:2002ij,fogli:2002au,deHolanda:2002iv})
one finds that the new data disfavour the high $\Delta m^2$ region,
which appears only at the 99.5\% \CL\ ($\Delta\chi^2
=10.7$).\footnote{Recently, an improved determination of the day/night
asymmetry in the 1496--day Super--Kamiokande data has been
released~\cite{SK-day/night-new}, which leads to a further
disfavouring of the high-LMA region.  Currently not enough information
is available to reproduce this result outside the Super--Kamiokande
collaboration.  Therefore, the day/night data is treated as in our
previous analyses~\cite{maltoni:2002ni}.}  The best fit point of the
global analysis occurs at
\begin{equation}\label{eq:bfp-sol+kaml}
\sin^2\theta_\Sol = 0.30 \,,\qquad
\Dms = 6.9\times 10^{-5}~\eVq \qquad\text{(solar+KamLAND data)}.
\end{equation}
We note that for the first time it is possible to obtain meaningful
bounds on solar neutrino parameters at the 5$\sigma$ level, showing
that neutrino physics enters the high precision age.
From the projections of the $\chi^2$ onto the $\Dms$ and
$\sin^2\theta_\Sol$ axes also shown in Fig.~\ref{fig:sol+kaml-region}
we find the following allowed ranges at 3$\sigma$ ($5\sigma$) for 1 \dof:
\begin{equation} \begin{split}\label{eq:solkam_ranges}
    0.23~(0.17) & \le \sin^2\theta_\Sol \le 0.39~(0.47)\,, \\
    5.4~(2.1)\times 10^{-5}~\eVq & \le \Dms  \le 9.4~(28)\times
    10^{-5}~\eVq\,. 
  \end{split}
\end{equation}

\section{Global three-neutrino analysis}
\label{sec:3nu}

\subsection{Notations}
\label{sec:notation}

In this section the three-neutrino oscillation parameters are
determined from a global analysis of the most recent neutrino
oscillation data. For earlier three-neutrino analyses see
Refs.~\cite{gonzalez-garcia:2000sq,fogli:2002au,Gonzalez-Garcia:2003qf}.
To fix the notation, we define the neutrino mass-squared differences
$\Dms \equiv \Delta m^2_{21} \equiv m^2_2 - m^2_1$ and $\Dma \equiv
\Delta m^2_{31} \equiv m^2_3 - m^2_1$, and use the standard
parameterization~\cite{hagiwara:2002fs,schechter:1980gr} for the
leptonic mixing matrix:
\begin{equation} \label{eq:mixing}
    U=\left(
    \begin{array}{ccc}
        c_{13} c_{12}
        & s_{12} c_{13}
        & s_{13} \\
        -s_{12} c_{23} - s_{23} s_{13} c_{12}
        & c_{23} c_{12} - s_{23} s_{13} s_{12}
        & s_{23} c_{13} \\
        s_{23} s_{12} - s_{13} c_{23} c_{12}
        & -s_{23} c_{12} - s_{13} s_{12} c_{23}
        & c_{23} c_{13}
    \end{array} \right) \,,
\end{equation}
where $c_{ij} \equiv \cos\theta_{ij}$ and $s_{ij} \equiv
\sin\theta_{ij}$. Furthermore, we use the notations $\theta_{12}
\equiv \theta_\Sol$ and $\theta_{23} \equiv \theta_\Atm$. Because of
the hierarchy $\Dms \ll \Dma$ it is a good approximation to set $\Dms
= 0$ in the analysis of atmospheric and K2K data\footnote{See
Ref.~\cite{Gonzalez-Garcia:2002mu} for a two-mass scale analysis of
atmospheric data.}, and to set $\Dma$ to infinity for the analysis of
solar and KamLAND data. This implies furthermore that the effect of a
possible Dirac CP-violating phase~\cite{schechter:1980gr} in the
lepton mixing matrix can be neglected~\footnote{The two Majorana
phases~\cite{schechter:1980gr} do not show up in oscillations but do
appear in lepton number violating processes~\cite{doi:1981yb}.}.  We
perform a general fit to the global data in the five-dimensional
parameter space $s^2_{12}, s^2_{23}, s^2_{13}, \Delta m^2_{21}, \Delta
m^2_{31}$, and show projections onto various one- or two-dimensional
sub-spaces.

\begin{figure}[t] \centering
 \includegraphics[width=.95\linewidth]{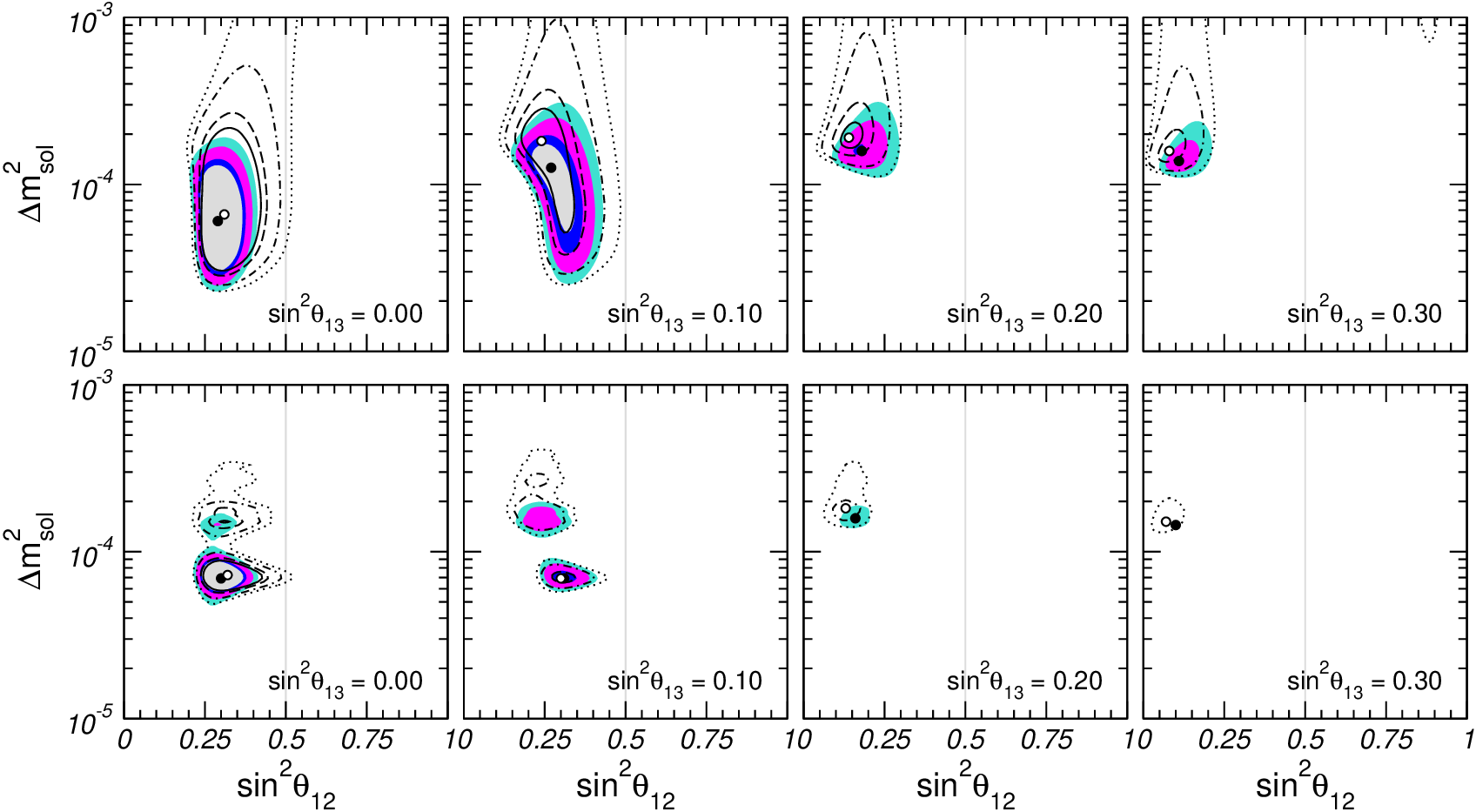}
 \caption{\label{fig:t13regions} Sections of the three-dimensional
   allowed regions in the ($\sin^2\theta_\Sol,\Dms$) plane at 90\%,
   95\%, 99\% and 3$\sigma$ \CL\ for 3 \dof\ for various
   $\sin^2\theta_{13}$ values from solar data (top) and solar+KamLAND
   data (bottom), before (lines) and after (colored regions) the
   SNO--salt data. The local minima in each plane after (before)
   SNO--salt data are marked by filled (open) dots.}
\end{figure}

\subsection{Three flavour solar neutrino oscillations}
\label{sec:sol3nu}

In this subsection we generalize the analysis of solar and KamLAND
data presented in Sec.~\ref{sec:solar+kaml} to three neutrino
flavours. Under the assumption of infinite $\Dma$ only the new
parameter $\theta_{13}$ appears for these data (see, \eg,
Ref.~\cite{gonzalez-garcia:2000sq}). In Fig.~\ref{fig:t13regions} we
show the results of a three parameter fit ($\sin^2\theta_\Sol, \Dms$,
$\sin^2\theta_{13}$) to solar and KamLAND data. Allowed regions are
shown for various values of $\sin^2\theta_{13}$ in the
($\sin^2\theta_\Sol, \Dms$) plane with respect to the global minimum,
which occurs for $\sin^2\theta_{13} = 0.02$ including SNO--salt data,
and for $\sin^2\theta_{13} = 0.01$ without the new data. Note that here
we calculate the allowed regions at a given confidence level for 3
\dof. From this figure one finds that the new SNO--salt data
contributes to the disappearance of allowed regions when
$\sin^2\theta_{13}$ increases. A comparison of the allowed regions
before and after the SNO--salt data shows that their size is
drastically reduced for all displayed values of $\sin^2\theta_{13}$.

\begin{figure}[t] \centering
\includegraphics[width=.5\linewidth]{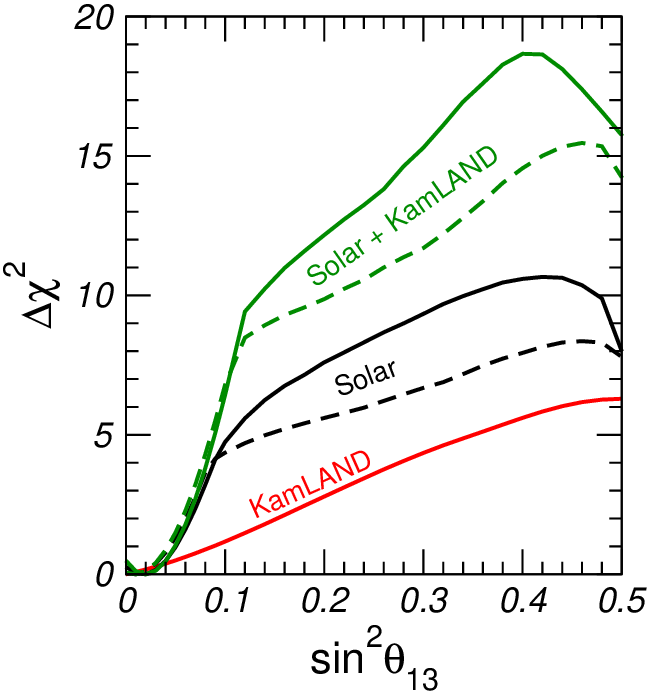}
    \caption{\label{fig:t13-solar}%
      $\Delta \chi^2$ profiles projected onto the $\sin^2\theta_{13}$
      axis, for solar and KamLAND data, before (dashed lines) and
      after (solid lines) the SNO-salt experiment.  }
\end{figure}

In Fig.~\ref{fig:t13-solar} we show the $\Delta\chi^2$ projected onto
the $\sin^2\theta_{13}$ axis for various data samples. From this
figure one can observe an improvement on the constraint on
$\sin^2\theta_{13}$ thanks to the new SNO--salt data for values of
$\sin^2\theta_{13} \gtrsim 0.08$. The shape of $\Delta\chi^2$ can be
understood from Fig.~\ref{fig:t13regions}: From the upper panels of
that figure one sees that increasing $\theta_{13}$ can be compensated
to some extent by increasing $\Dms$. Since the new SNO--salt data
disfavours large values of $\Dms$ the bound improves. Also KamLAND
data acts in a similar way. Since the minimum around $\Dms \sim
7\times 10^{-5}$~eV$^2$ is preferred the ``jump'' of the local minimum
into the high-LMA-MSW region, which is visible in the lower panels of
Fig.~\ref{fig:t13regions} and which leads to the kink in the
$\Delta\chi^2$ shown in Fig.~\ref{fig:t13-solar}, occurs at larger
values of $\sin^2\theta_{13}$. We have also verified explicitly that
assumptions regarding the statistical treatment of the solar boron
flux have a rather small effect on these results. Finally we note
that, as will be seen in Subsec.~\ref{sec:subleading}, the bound
resulting from CHOOZ and atmospheric data still dominates the overall
constraint on $\sin^2\theta_{13}$ in the global analysis.

\subsection{Global analysis of solar, atmospheric, reactor and
      accelerator data} 
\label{sec:global}

\begin{table}[t] \centering
    \catcode`?=\active \def?{\hphantom{0}}
    \begin{tabular}{|@{\quad}>{\rule[-2mm]{0pt}{6mm}}l@{\quad}|@{\quad}c@{\quad}|@{\quad}c@{\quad}|@{\quad}c@{\quad}|@{\quad}c@{\quad}|}
        \hline
        parameter & best fit & 2$\sigma$ & 3$\sigma$ & 5$\sigma$
        \\
        \hline
        $\Delta m^2_{21}\: [10^{-5}\eVq]$ & 6.9?? & 6.0--8.4 & 5.4--9.5 & ??2.1--28?? \\
        $\Delta m^2_{31}\: [10^{-3}\eVq]$ & 2.6?? & 1.8--3.3 & 1.4--3.7 & 0.77--4.8? \\
        $\sin^2\theta_{12}$ & 0.30? & 0.25--0.36 & 0.23--0.39 & 0.17--0.48 \\ 
        $\sin^2\theta_{23}$ & 0.52? & 0.36--0.67 & 0.31--0.72 & 0.22--0.81 \\ 
        $\sin^2\theta_{13}$ & 0.006 & $\leq$ 0.035 & $\leq$ 0.054 & $\leq$ 0.11 \\
        \hline
    \end{tabular}
    \caption{ \label{tab:summary} Best-fit values, 2$\sigma$, 
      3$\sigma$ and 5$\sigma$ intervals (1 \dof) for the three-flavour neutrino
      oscillation parameters from global data including solar,
      atmospheric, reactor (KamLAND and CHOOZ) and accelerator
      (K2K) experiments.}
\end{table}

In addition to the solar and KamLAND data described so far we now
include all the charged-current atmospheric neutrino data, such as the
most recent ones of the
Super--Kamiokande~\cite{fukuda:1998mi,shiozawa:2002} and
MACRO~\cite{surdo:2002rk} experiments.  The Super--Kamiokande data
include the $e$-like and $\mu$-like data samples of sub- and multi-GeV
contained events (10 bins in zenith angle), as well as the stopping (5
angular bins) and through-going (10 angular bins) up-going muon data
events.  As previously, we do not use the information on $\nu_\tau$
appearance, multi-ring $\mu$ and neutral-current events since an
efficient Monte-Carlo simulation of these data would require a more
detailed knowledge of the Super--Kamiokande experiment, and in
particular of the way the neutral-current signal is extracted from the
data.  From MACRO we use the through-going muon sample divided in 10
angular bins~\cite{surdo:2002rk}. For details of our analysis see
Refs.~\cite{gonzalez-garcia:2000sq,maltoni:2002ni} and references
therein. Furthermore, we include the first spectral data from the K2K
long-baseline accelerator experiment~\cite{ahn:2002up}. We use the 29
single-ring muon events, grouped into 6 energy bins, and we perform a
spectral analysis similar to the one described in
Ref.~\cite{Fogli:2003th}.  Finally, we take into account in our fit
the constraints from the CHOOZ reactor
experiment~\cite{apollonio:1999ae}. The best fit values
and the allowed intervals of the parameters $s^2_{12}, s^2_{23},
s^2_{13}, \Delta m^2_{21}, \Delta m^2_{31}$ from the global data are
given in Tab.~\ref{tab:summary}. This table summarizes the current
status of the three flavour neutrino oscillation parameters.

\begin{figure}[t] \centering
\includegraphics[width=.6\linewidth]{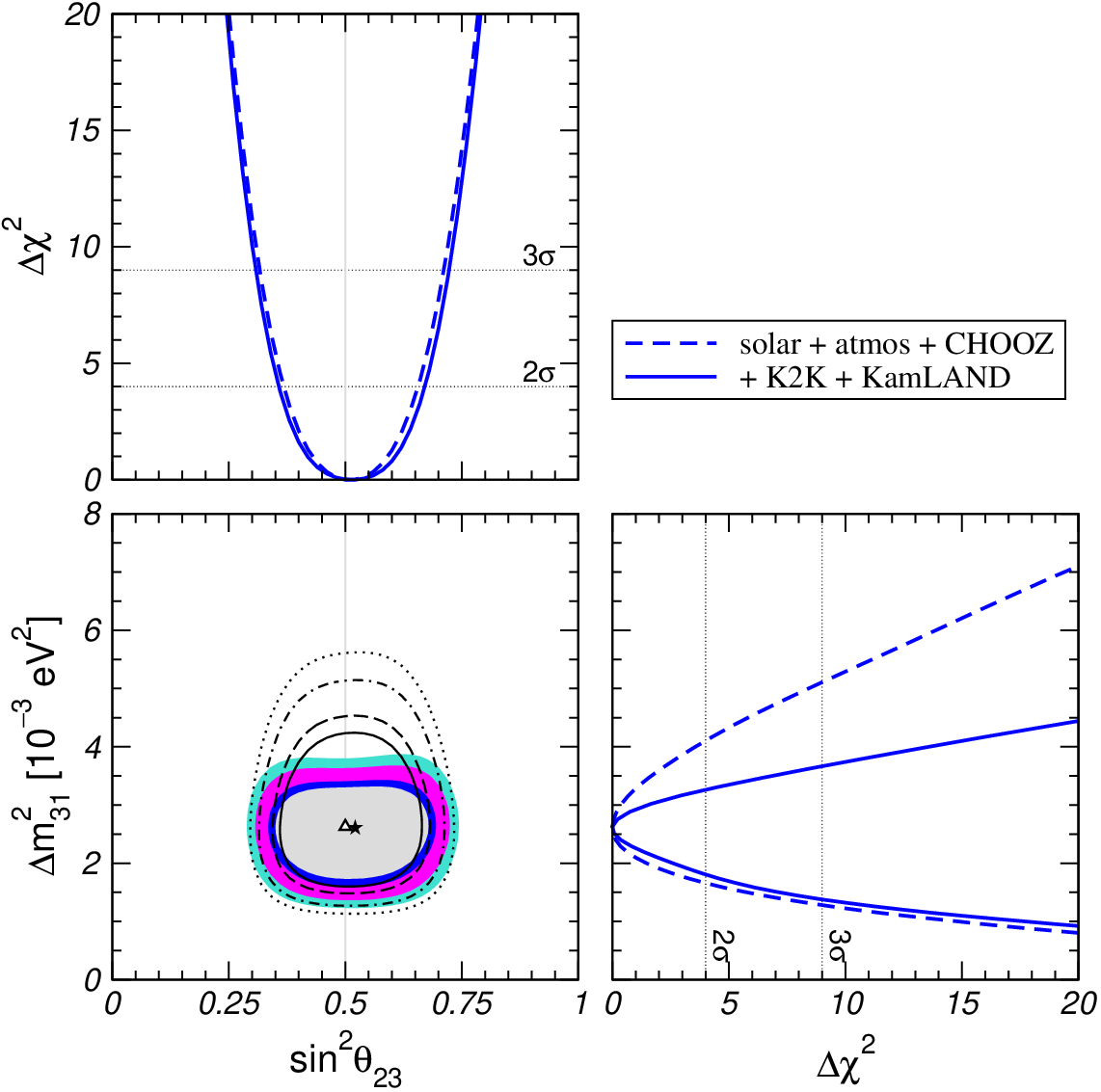}
    \caption{\label{fig:atm+k2k} Projections of the allowed regions at
      90\%, 95\%, 99\%, and 3$\sigma$ \CL\ for 2 \dof\ onto the plane
      of $\sin^2\theta_\Atm\equiv \sin^2\theta_{23}$ and $\Dma \equiv
      \Delta m^2_{31}$. The regions delimited by the lines correspond
      to atmospheric+solar+CHOOZ data, for the colored regions also
      K2K and KamLAND data is added. Also shown is the $\Delta \chi^2$
      as a function of $\sin^2\theta_\Atm$ and $\Dma$, minimized with
      respect to all un-displayed parameters.}
\end{figure}

For completeness we show in Fig.~\ref{fig:atm+k2k} the projection of
the allowed regions from the global fit onto the plane of the
atmospheric neutrino parameters. We find that the first 29 events from
K2K included here start already to constrain the upper region of
$\Dma$.  We note that recently the Super--Kamiokande collaboration has
presented a preliminary reanalysis of their atmospheric
data~\cite{sk:aachen}.  The up-date includes changes in the detector
simulation, data analysis and input atmospheric neutrino fluxes. These
changes lead to a slight downward shift of the mass-splitting to the
best fit value of $\Dma = 2\times 10^{-3}~\eVq$. Currently it is not
possible to recover enough information from Ref.~\cite{sk:aachen} to
incorporate the corresponding changes in our codes. We note, however,
that the quoted value for $\Dma$ is statistically compatible with our
result. For $\Dma = 2\times 10^{-3}~\eVq$ and maximal mixing we obtain
a $\Delta \chi^2 = 1.3$.

\subsection{Status of $\theta_{13}$ and the mass hierarchy parameter
    $\alpha = \Dms/\Dma$}
\label{sec:subleading}

In this subsection we discuss in detail the current information on the
small parameters relevant for sub-leading oscillation effects.  First
we consider the mixing angle $\theta_{13}$, which at the moment is the
last unknown angle in the three-neutrino mixing matrix. Only an upper
bound exists, which used to be dominated by the CHOOZ reactor
experiment~\cite{apollonio:1999ae}. Currently a large effort is put to
determine this angle in future experiments (see, \eg,
Refs.~\cite{pakvasa:2003zv,Huber:2003pm,Huber:2002mx}). 

\begin{figure}[t] \centering
\includegraphics[width=.5\linewidth]{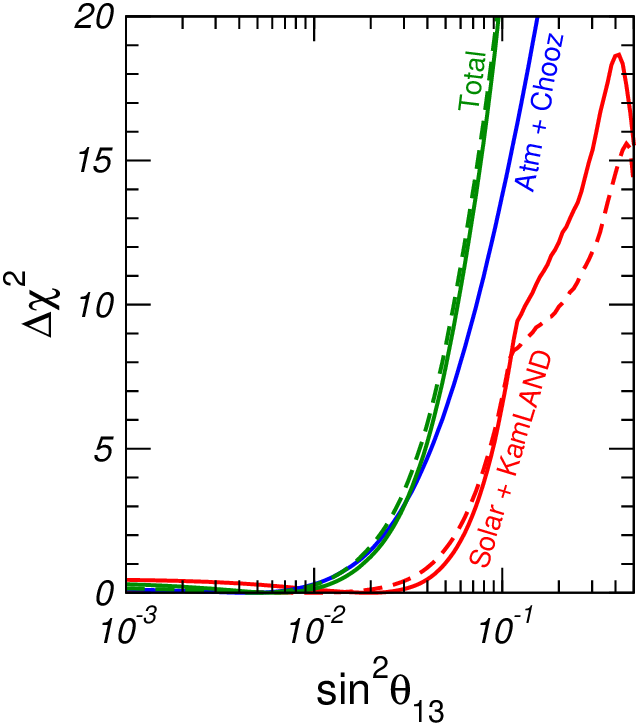}
    \caption{\label{fig:t13-compare}%
      $\Delta \chi^2$ profiles projected onto the $\sin^2\theta_{13}$
      axis, for solar+KamLAND, atmospheric+CHOOZ, and for the global
      data before (dashed lines) and after (solid lines) the SNO--salt
      experiment.  }
\end{figure}

In Fig.~\ref{fig:t13-compare} we show the $\Delta\chi^2$ as a function
of $\sin^2\theta_{13}$ for different data sample choices. One can see
how the bound on $\sin^2\theta_{13}$ as implied by the CHOOZ
experiment in combination with the atmospheric neutrino data still
provides the main restriction on $\sin^2\theta_{13}$. We find the
following bounds at 90\% \CL\ (3$\sigma$) for 1 \dof:
\begin{equation}\label{eq:th13}
    \sin^2\theta_{13} \le \left\{ \begin{array}{l@{\qquad}l}
      0.070~(0.12) & \text{(solar+KamLAND)} \\
      0.028~(0.066) & \text{(CHOOZ+atmospheric)} \\
      0.029~(0.054) & \text{(global data)}
  \end{array} \right.
\end{equation}

\begin{figure}[t] \centering
\includegraphics[width=.6\linewidth]{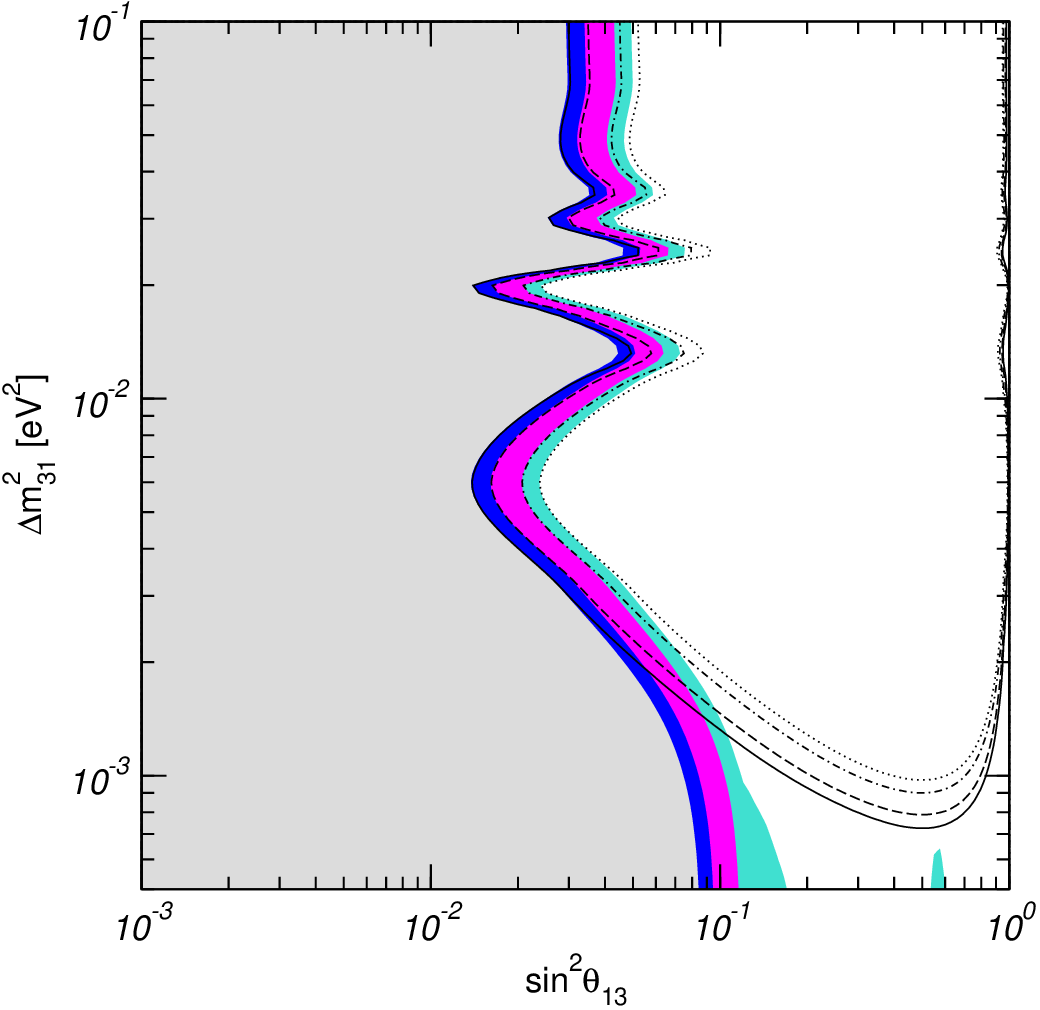}
    \caption{\label{fig:t13-solar-chooz}%
    Allowed regions in the ($\sin^2\theta_{13}, \Dma$) plane at 
    90\%, 95\%, 99\%, and 3$\sigma$ from CHOOZ data alone (lines) and
    CHOOZ+solar+KamLAND data (colored regions).}
\end{figure}

However, we note that the solar data contributes in an important way
to the constraint on $\sin^2\theta_{13}$ for lower values of $\Dma$.
In particular, the down-ward shift of $\Dma$ reported in
Ref.~\cite{sk:aachen} implies a significant loosening of the CHOOZ
bound on $\sin^2\theta_{13}$, since this bound gets quickly weak when
$\Dma$ decreases (see, \eg, Ref.~\cite{Fogli:2003am}). Such loosening
in sensitivity is prevented to some extent by solar neutrino data. In
Fig.~\ref{fig:t13-solar-chooz} we show the allowed regions in the
($\sin^2\theta_{13}, \Dma$) plane from an analysis including solar and
reactor neutrino data (CHOOZ and KamLAND). One finds that, although
for larger $\Dma$ values the bound on $\sin^2\theta_{13}$ is dominated
by the Chooz + atmospheric data, for low $\Dma = 10^{-2} \eVq$ the
solar + KamLAND bound is comparable to that coming from Chooz +
atmospheric.

For example, fixing $\Dma = 2\times 10^{-3}~\eVq$ we obtain at
90\% \CL\ (3$\sigma$) for 1 \dof\ the bound
\begin{equation}
  \sin^2\theta_{13} \le 0.035~(0.066) \qquad (\Dma = 2\times 10^{-3}~\eVq) \,,
\end{equation}
which is only marginally worse than the bound from global data shown
in Eq.~(\ref{eq:th13}).  

\begin{figure}[t] \centering
\includegraphics[width=.5\linewidth]{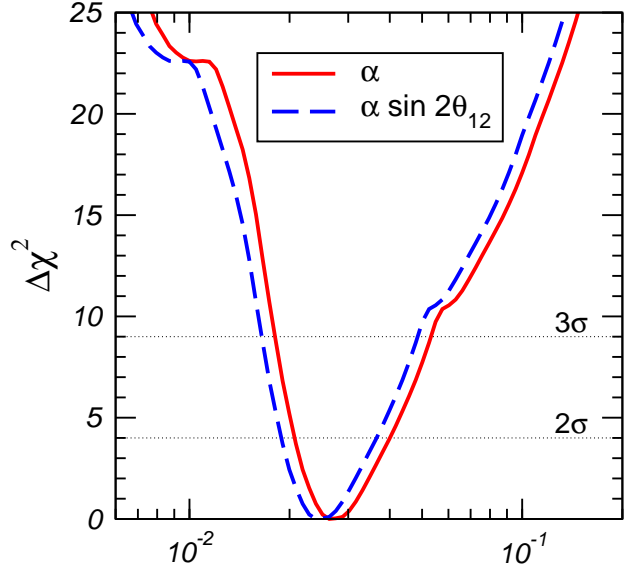}
    \caption{\label{fig:alpha}%
      $\Delta \chi^2$ from global oscillation data as a function of 
      $\alpha \equiv \Dms / \Dma$ and $\alpha\sin 2\theta_{12}$.}
\end{figure}

For the exploration of genuine three-flavour effects such as
CP-violation the mass hierarchy parameter $\alpha \equiv \Dms/\Dma$ is
of crucial importance since, in a three--neutrino scheme, CP violation
disappears in the limit where two neutrinos become
degenerate~\cite{Schechter:1979bn}.  Therefore we show in
Fig.~\ref{fig:alpha} the $\Delta\chi^2$ from the global data as a
function of this parameter. Also shown in this figure is the
$\Delta\chi^2$ as a function of the parameter combination $\alpha \sin
2\theta_{12}$, since to leading order in the long baseline
$\nu_e\to\nu_\mu$ oscillation probability solar parameters appear in
this particular combination~\cite{Freund:2001pn}. We obtain the
following best fit values and 3$\sigma$ intervals:
\begin{equation}
\begin{split}
   \alpha = 0.026\,,\quad & 0.018 \le \alpha \le 0.053\,,\\
   \alpha \sin 2\theta_{12} = 0.024\,,\quad & 
    0.016 \le \alpha \sin 2\theta_{12} \le 0.049\,.
\end{split}
\end{equation}

\section{Conclusions}
\label{sec:conclusions}

\begin{figure}[t] \centering
\includegraphics[width=.9\linewidth]{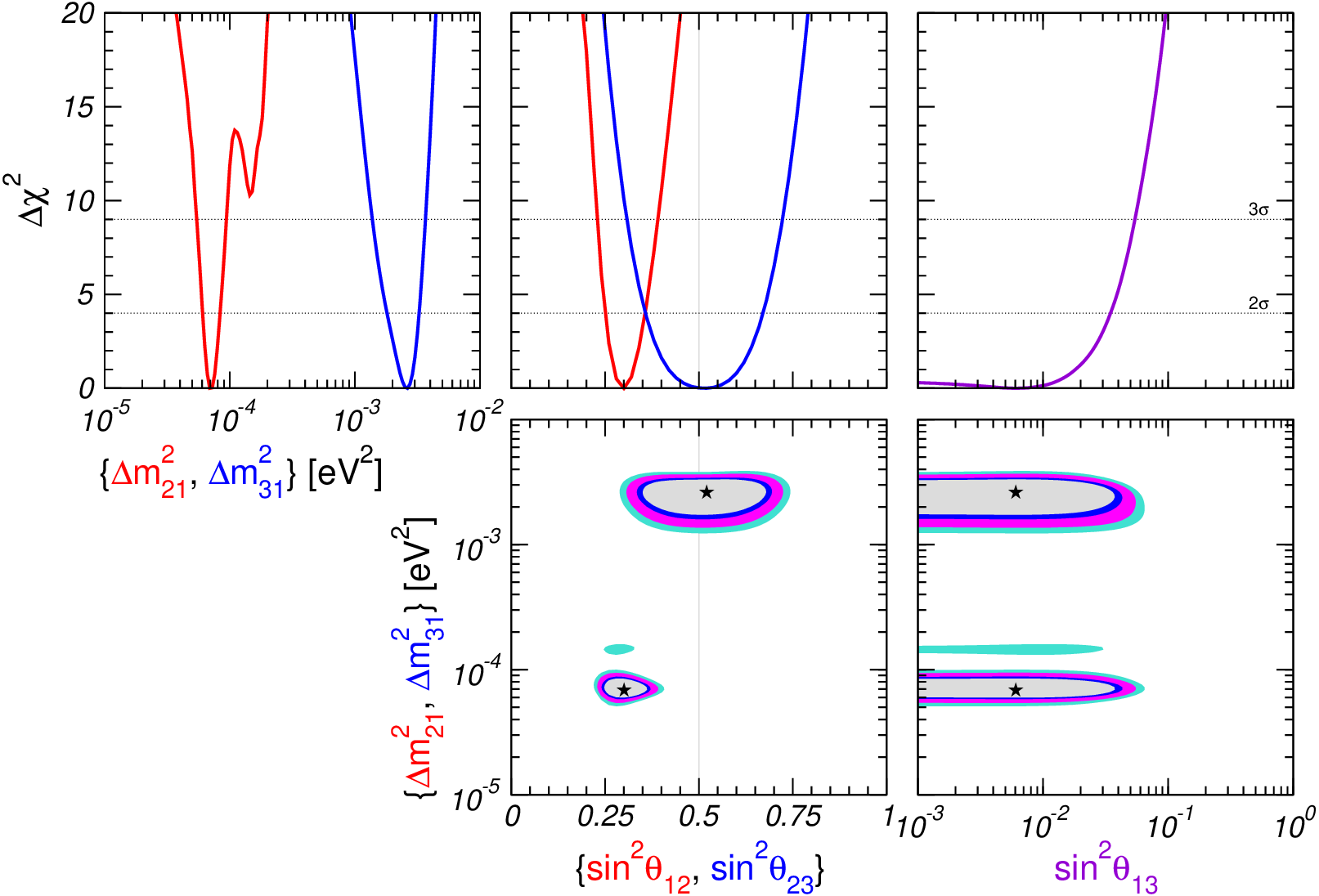}
    \caption{\label{fig:global} Projections of the allowed regions
      from the global oscillation data at 90\%, 95\%, 99\%, and
      3$\sigma$ \CL\ for 2 \dof\ for various parameter
      combinations. Also shown is $\Delta \chi^2$ as a function of the
      oscillation parameters $\sin^2\theta_{12}, \sin^2\theta_{23},
      \sin^2\theta_{13}, \Delta m^2_{21}, \Delta m^2_{31}$, minimized
      with respect to all un-displayed parameters.}
\end{figure}

We have performed a global analysis of neutrino oscillation data in
the three--neutrino scheme, including the recent improved measurement
of the neutral current events at SNO. We include in our fit all
current solar neutrino data, reactor neutrino data from KamLAND and
CHOOZ, the atmospheric neutrino data from Super--Kamiokande and MACRO,
as well as the first data from the K2K long-baseline accelerator
experiment. We have discussed the implications of the improved neutral
current measurement on the determination of the solar neutrino
oscillation parameters. The SNO--salt results reject the previously
allowed high-mass branch of $\Dms$ at about 3$\sigma$. Moreover, it
rules out maximal solar mixing at more than 5$\sigma$, precluding the
possibility of bi--maximal models of neutrino mixing.

Furthermore, we have determined the current best fit values and
allowed ranges for the three--flavour oscillation parameters from the
global oscillation data. Our three--neutrino results are summarized in
Tab.~\ref{tab:summary} and in Fig.~\ref{fig:global}, showing the
allowed regions and $\chi^2$ projections for the five oscillation
parameters
$\sin^2\theta_\Sol,\sin^2\theta_\Atm,\sin^2\theta_{13},\Dms,\Dma$.  In
addition we have discussed in detail the limits on the small mixing
angle $\sin^2\theta_{13}$, as well as the hierarchy parameter $\alpha
= \Dms/\Dma$. These small parameters are relevant for genuine three
flavour effects, and restrict the magnitude of leptonic CP violation
that one may potentially probe at future experiments like super beams
or neutrino factories. In particular, we have seen how the improvement
on the $\sin^2\theta_{13}$ limit that follows from the new SNO--salt
data solar neutrino experiments can not yet match the sensitivity on
$\sin^2\theta_{13}$ reached at reactor experiments. However, the solar
data do play am important role in stabilizing the constraint on
$\sin^2\theta_{13}$ with respect to variation of $\Dma$.  For small
enough $\Dma$ values the solar data probe $\sin^2\theta_{13}$ at a
level comparable to that of the current reactor experiments.

\acknowledgments

This work was supported by Spanish grant BFM2002-00345, by the
  European Commission RTN network HPRN-CT-2000-00148 and by the
  European Science Foundation network grant N.~86. M.M.\ is supported
  by contract HPMF-CT-2000-01008 and M.A.T.\ is supported by the
  M.E.C.D.\ fellowship AP2000-1953. T.S.\ is supported by the
  ``Sonderforschungsbereich 375-95 f{\"u}r Astro-Teilchenphysik'' der
  Deutschen Forschungsgemeinschaft.


\begin{thebibliography}{10}
  
\bibitem{ahmad:2003} SNO collaboration, ``Measurement of the Total
  Active 8B Solar Neutrino Flux at the Sudbury Neutrino Observatory
  with Enhanced Neutral Current Sensitivity'', Q.~R. Ahmad {\em
    et~al.}, \newblock [nucl-ex/0309004].

\bibitem{eguchi:2002dm}
KamLAND collaboration, K.~Eguchi {\em et~al.},
\newblock Phys. Rev. Lett. {\bf 90}, 021802 (2003), [hep-ex/0212021].

\bibitem{cleveland:1998nv}
B.~T. Cleveland {\em et~al.},
\newblock Astrophys. J. {\bf 496}, 505 (1998).

\bibitem{davis:1994jw}
R.~Davis,
\newblock Prog. Part. Nucl. Phys. {\bf 32}, 13 (1994).

\bibitem{abdurashitov:2002nt}
SAGE collaboration, J.~N. Abdurashitov {\em et~al.},
\newblock J. Exp. Theor. Phys. {\bf 95}, 181 (2002), [astro-ph/0204245].

\bibitem{abdurashitov:1999zd}
SAGE collaboration, J.~N. Abdurashitov {\em et~al.},
\newblock Phys. Rev. {\bf C60}, 055801 (1999), [astro-ph/9907113].

\bibitem{hampel:1998xg}
GALLEX collaboration, W.~Hampel {\em et~al.},
\newblock Phys. Lett. {\bf B447}, 127 (1999).

\bibitem{altmann:2000ft}
GNO collaboration, M.~Altmann {\em et~al.},
\newblock Phys. Lett. {\bf B490}, 16 (2000), [hep-ex/0006034].

\bibitem{cattadori:2002rd}
GNO collaboration, C.~M. Cattadori,
\newblock Nucl. Phys. Proc. Suppl. {\bf 110}, 311 (2002).

\bibitem{fukuda:2002pe}
Super--Kamiokande collaboration, S.~Fukuda {\em et~al.},
\newblock Phys. Lett. {\bf B539}, 179 (2002), [hep-ex/0205075].

\bibitem{ahmad:2002jz}
SNO collaboration, Q.~R. Ahmad {\em et~al.},
\newblock Phys. Rev. Lett. {\bf 89}, 011301 (2002), [nucl-ex/0204008].

\bibitem{ahmad:2002ka}
SNO collaboration, Q.~R. Ahmad {\em et~al.},
\newblock Phys. Rev. Lett. {\bf 89}, 011302 (2002), [nucl-ex/0204009].

\bibitem{ahmad:2001an}
SNO collaboration, Q.~R. Ahmad {\em et~al.},
\newblock Phys. Rev. Lett. {\bf 87}, 071301 (2001), [nucl-ex/0106015].

\bibitem{guzzo:2001mi}
M.~Guzzo {\em et~al.},
\newblock Nucl. Phys. {\bf B629}, 479 (2002), [hep-ph/0112310 v3
  KamLAND-updated version].

\bibitem{barranco:2002te}
J.~Barranco, O.~G. Miranda, T.~I. Rashba, V.~B. Semikoz and J.~W.~F. Valle,
\newblock Phys. Rev. {\bf D66}, 093009 (2002), [hep-ph/0207326 v3
  KamLAND-updated version].

\bibitem{maltoni:2002aw}
M.~Maltoni, T.~Schwetz and J.~W.~F. Valle,
\newblock Phys. Rev. {\bf D67}, 093003 (2003), [hep-ph/0212129].

\bibitem{Bahcall:2002ij}
J.~N. Bahcall, M.~C. Gonzalez-Garcia and C.~Pena-Garay,
\newblock hep-ph/0212147.

\bibitem{fogli:2002au}
G.~L. Fogli {\em et~al.},
\newblock Phys. Rev. {\bf D67}, 073002 (2003), [hep-ph/0212127].

\bibitem{deHolanda:2002iv}
P.~C.~de Holanda and A.~Y.~Smirnov,
\newblock JCAP {\bf 0302}, 001 (2003) [hep-ph/0212270].

\bibitem{wolfenstein:1978ue}
L.~Wolfenstein,
\newblock Phys. Rev. {\bf D17}, 2369 (1978).

\bibitem{mikheev:1985gs}
S.~P. Mikheev and A.~Y. Smirnov,
\newblock Sov. J. Nucl. Phys. {\bf 42}, 913 (1985).

\bibitem{fukuda:1994mc}
Kamiokande collaboration, Y.~Fukuda {\em et~al.},
\newblock Phys. Lett. {\bf B335}, 237 (1994).

\bibitem{becker-szendy:1995vr}
IMB collaboration, R.~Becker-Szendy {\em et~al.},
\newblock Nucl. Phys. Proc. Suppl. {\bf 38}, 331 (1995).

\bibitem{allison:1999ms}
Soudan-2 collaboration, W.~W.~M. Allison {\em et~al.},
\newblock Phys. Lett. {\bf B449}, 137 (1999), [hep-ex/9901024].

\bibitem{shiozawa:2002}
M.~Shiozawa,
\newblock (2002),
\newblock Talk at Neutrino 2002, http://neutrino2002.ph.tum.de/.

\bibitem{fukuda:1998mi}
Super--Kamiokande collaboration, Y.~Fukuda {\em et~al.},
\newblock Phys. Rev. Lett. {\bf 81}, 1562 (1998), [hep-ex/9807003].

\bibitem{surdo:2002rk}
MACRO collaboration, A.~Surdo,
\newblock Nucl. Phys. Proc. Suppl. {\bf 110}, 342 (2002).

\bibitem{ahn:2002up}
K2K collaboration, M.~H. Ahn {\em et~al.},
\newblock Phys. Rev. Lett. {\bf 90}, 041801 (2003), [hep-ex/0212007].

\bibitem{apollonio:1999ae}
CHOOZ collaboration, M.~Apollonio {\em et~al.},
\newblock Phys. Lett. {\bf B466}, 415 (1999), [hep-ex/9907037].

\bibitem{boehm:2001ik}
F.~Boehm {\em et~al.},
\newblock Phys. Rev. {\bf D64}, 112001 (2001), [hep-ex/0107009].

\bibitem{maltoni:2002ni}
M.~Maltoni, T.~Schwetz, M.~A. Tortola and J.~W.~F. Valle,
\newblock Phys. Rev. {\bf D67}, 013011 (2003), [hep-ph/0207227 v3
  KamLAND-updated version].

\bibitem{ahmad:2003a} SNO HOWTO kit, Q.~R. Ahmad {\em et~al.},
  \newblock [http://www.sno.phy.queensu.ca/sno/].

\bibitem{taup} H.~Robertson, Talk at the TAUP03 conference, 
September 5--9, 2003, Seattle, Washington, 
[http://mocha.phys.washington.edu/taup2003/].

\bibitem{gonzalez-garcia:2000sq}
M.~Gonzalez-Garcia, M.~Maltoni, C.~Pena-Garay and J.~W.~F. Valle,
\newblock Phys. Rev. {\bf D63}, 033005 (2001), [hep-ph/0009350].

\bibitem{Fogli:2002pt}
G.~L.~Fogli, E.~Lisi, A.~Marrone, D.~Montanino and A.~Palazzo,
\newblock Phys.\ Rev.\ D {\bf 66}, 053010 (2002) [hep-ph/0206162].

\bibitem{Balantekin:2003jm}
A.~B.~Balantekin and H.~Yuksel, hep-ph/0309079.
 
\bibitem{bahcall:1998wm}
J.~N. Bahcall, S.~Basu and M.~H. Pinsonneault,
\newblock Phys. Lett. {\bf B433}, 1 (1998), [astro-ph/9805135].
%
J.~N.~Bahcall, M.~H.~Pinsonneault and S.~Basu,
\newblock Astrophys.\ J.\  {\bf 555}, 990 (2001) [astro-ph/0010346].

\bibitem{pakvasa:2003zv} For recent reviews see S.~Pakvasa and
  J.~W.~F. Valle, \newblock hep-ph/0301061, \newblock Prepared for
  Special Issue of Proceedings of the Indian National Academy of
  Sciences on Neutrinos; 
%
V.~Barger, D.~Marfatia and K.~Whisnant,
hep-ph/0308123, and references therein.

\bibitem{schwetz:2003se}
T.~Schwetz,
\newblock hep-ph/0308003.

\bibitem{SK-day/night-new}
M.B.~Smy \EtAl, Super--Kamiokande Coll., hep-ex/0309011.

\bibitem{Gonzalez-Garcia:2003qf}
M.~C. Gonzalez-Garcia and C.~Pena-Garay,
\newblock hep-ph/0306001.

\bibitem{hagiwara:2002fs}
Particle Data Group, K.~Hagiwara {\em et~al.},
\newblock Phys. Rev. {\bf D66}, 010001 (2002).

\bibitem{schechter:1980gr}
J.~Schechter and J.~W.~F. Valle,
\newblock Phys. Rev. {\bf D22}, 2227 (1980).

\bibitem{Gonzalez-Garcia:2002mu}
M.~C. Gonzalez-Garcia and M.~Maltoni,
\newblock Eur. Phys. J. {\bf C26}, 417 (2003), [hep-ph/0202218].

\bibitem{doi:1981yb}
M.~Doi, T.~Kotani, H.~Nishiura, K.~Okuda and E.~Takasugi,
\newblock Phys. Lett. {\bf B102}, 323 (1981).
J.~Schechter and J.~W.~Valle,
\newblock Phys.\ Rev.\ D {\bf 23}, 1666 (1981) and
Phys.\ Rev.\ D {\bf 24}, 1883 (1981) [Err-ibid.\ D {\bf 25}, 283
(1982)].

\bibitem{Fogli:2003th}
G.~L. Fogli, E.~Lisi, A.~Marrone and D.~Montanino,
\newblock Phys. Rev. {\bf D67}, 093006 (2003), [hep-ph/0303064].

\bibitem{sk:aachen}
Y. Hayato, Super--Kamiokande Coll., talk at the HEP2003 conference (Aachen,
Germany, 2003), http://eps2003.physik.rwth-aachen.de

\bibitem{Huber:2003pm}
P.~Huber, M.~Lindner, T.~Schwetz and W.~Winter,
Nucl.\ Phys.\ B {\bf 665}, 487 (2003) [hep-ph/0303232].

\bibitem{Huber:2002mx}
P.~Huber, M.~Lindner and W.~Winter,
Nucl.\ Phys.\ B {\bf 645}, 3 (2002) [hep-ph/0204352];
M.~Apollonio {\it et al.},
hep-ph/0210192;
M.~M.~Alsharoa {\it et al.}  [Muon Collider/Neutrino Factory Collaboration],
hep-ex/0207031.

\bibitem{Fogli:2003am}
G.~L.~Fogli, E.~Lisi, A.~Marrone, D.~Montanino, A.~Palazzo and A.~M.~Rotunno,
hep-ph/0308055.

\bibitem{Schechter:1979bn}
J.~Schechter and J.~W.~Valle,
Phys.\ Rev.\ D {\bf 21}, 309 (1980).

\bibitem{Freund:2001pn}
M.~Freund,
Phys.\ Rev.\ D {\bf 64}, 053003 (2001) [hep-ph/0103300].

\end{thebibliography}
\end{document}